\documentclass[10pt]{article}
\usepackage[OE]{express}

\begin{document}
\title{Single shot large field of view imaging with scattering media by spatial demultiplexing}

\author{Sujit Kumar Sahoo,\authormark{1,2,*} Dongliang Tang,\authormark{1} and Cuong Dang\authormark{1,*}}

\address{\authormark{1}Centre for OptoElectronics and Biophotonics (COEB), School of Electrical and Electronic Engineering, The Photonic Institute (TPI), Nanyang Technological University Singapore, 50 Nanyang Avenue, 639798, Singapore\\
\authormark{2}Department of Statistics and Applied Probability, National University of Singapore, 117546, Singapore}

\email{\authormark{*} sujit@pmail.ntu.edu.sg, hcdang@ntu.edu.sg} 



\begin{abstract*}
Optically focusing and imaging through strongly scattering media are challenging tasks but have widespread applications from scientific research to biomedical applications and daily life. Benefiting from the memory effect (ME) for speckle intensity correlations, only one single-shot speckle pattern can be used for the high quality recovery of the objects and avoiding some complicated procedures to reduce scattering effects. In spite of all the spatial information from a large object being embedded in a single speckle image, ME gives a strict limitation to the field of view (FOV) for imaging through scattering media. Objects beyond the ME region cannot be recovered and only produce unwanted speckle patterns, causing reduction in the speckle contrast and recovery quality. Here, we extract the spatial information by utilizing these unavoidable speckle patterns, and enlarge the FOV of the optical imaging system. Regional point spreading functions (PSFs), which are fixed and only need to be recorded once for all time use, are employed to recover corresponding spatial regions of an object by deconvolution algorithm. Then an automatic weighted averaging in an iterative process is performed to obtain the object with significantly enlarged FOV. Our results present an important step toward an advanced imaging technique with strongly scattering media.
\end{abstract*}



\section{Introduction}
Optical imaging is one of the most simple and straightforward ways to recognize the surrounding world with large amount of information. The inventions and developments of the telescope, microscope and camera technology have driven innovations in numerous fields from the astronomical observation\cite{Noll76, Tyson15} to biomedical research\cite{Gabri96, Ntzia10}, also in our daily life. Unfortunately, there are still some real-life obstacles such as the atmospheric turbulence, biological tissues or even bead curtains, which degrade the visibility of the interested objects at different levels. These scattering media contain inhomogeneity of refractive index or absorption coefficient to generate random phase or amplitude distortions on the ideal light paths from the original objects. This results in complex speckle patterns on detections and impairs the optical observations. Thus, imaging through those scattering media has become a hot topic and an important technical challenge in the past decades.

To overcome these limitations, various novel, breakthrough and practical approaches have been proposed with the advances of the optically scattering theory and optoelectronic devices. One of the successful techniques, optical coherence tomography (OCT), can utilize the un-scattered ballistic light and a scanning gate window to capture the static images, conversely the required amount of the ballistic photons will give a limitation of the penetration depth through scattering media\cite{Huang91, Nasr03}. As the scattering media mainly scramble the phase information after light propagation in most cases, another straightforward method is trying to compensate the phase perturbation. Learning from the adaptive optics in the astronomical observation\cite{Tyson15}, one can use a spatial light modulator (SLM) to play the role of the deformable mirrors and reduce the aberrations\cite{Velle10,Katz12,He13,Horst15}. An iterative algorithm with the assistance of a bright guiding star as a part of the feedback mechanism effectively reduces the aberrations, paving the way for focusing and imaging through scattering media. By recording the transmitted wavefront of the scattered light in an optical setup with strict alignment, one can generate a phase-conjugate wavefront using SLM\cite{Cui09,Xu11,Si12,Velle12,Judke13}. 

Another set of ideas was proposed by exploiting the speckle intensity correlation for imaging, thank to ME\cite{Freun88,Freun90,Berto12,Katz14,Edrei16,Porat16,Singh17,Shi17,Wu16,Edrei16r,Zhuan16,Cua17}.  The ME states that the random speckle patterns observed after the scattering medium are shifted (i.e. highly correlated) when the illumination angle varies within a certain range. By estimating this ME region, a very sparse object of size smaller than this can be recovered noninvasively with an ultra narrow-band illumination. Because of the ME and randomness of the speckle pattern, the autocorrelation of the object mostly preserves through scattering media. A single-shot speckle pattern is enough to reconstruct the object with the help of a phase retrieval algorithm\cite{Katz14,Edrei16,Porat16,Wu16,Singh17,Shi17}. The ME region on the object plane is inversely proportional to the thickness $L$ of the scattering medium and proportional to the distance $u$ from the object plane to the scattering medium, according to the formula: ${u\lambda}/{\pi L}$ \cite{Freun88,Freun90}. Objects larger than the ME region produce unwanted speckle patterns, the object autocorrelation can not be preserved, and the phase retrieval is not possible anymore. Deconvolution imaging becomes useful in this case, where the scattering medium could be treated as a scattering lens \cite{Edrei16r,Zhuan16}. One can make full use of the intrinsic ME region by suitably positioning the scattering medium and optics components \cite{Zhuan16}. However, the FOV of the optical imaging systems is still limited by the ME region of strongly scattering media, and unwanted speckle patterns only result in the reconstruction artifacts. 

In this work, we demonstrate a deconvolution imaging technique to produce a larger FOV from a single-shot image with scattering media. Regional point spreading functions (PSFs), which are fixed and only need to be recorded once for all time use, are employed to recover corresponding spatial regions of an object by deconvolution algorithm. Then an automatic weighted averaging in an iterative process is performed to obtain the object with significantly enlarged FOV. We make use of those uncorrelated speckle patterns generated by objects beyond the ME region, which are previously considered as unwanted noise. The motivation is to maximize the utilization of the information captured by the imager, which is the superposition of multiple speckle patterns coming from the different regions of an extra-large object. Our results present an important step toward an advanced imaging technique with strongly scattering media. The detailed principle is described in the following section.

\section{Principle}
The object is simply a composition of point sources with various positions and intensities. For a linear optical imaging system, an image is expressed as a superposition of the multiple PSFs from different spatial positions with corresponding intensity. The following equation is a mathematical representation of the image formation.
\begin{equation}
\label{eq1}
I(x,y)=\sum_{(i,j)}{O(i,j)PSF_{ij}(x-i,y-j)}
\end{equation}
where $O$ is the object, and $PSF_{ij}$ is the optical response function of a point source located at position $(i,j)$. In an ideal scenario, the image becomes identical to the object $I=O$ when these PSFs are identical impulse functions, which denotes an ideal optical imaging system. In contrast, the scattering medium produces not only random but also variable PSFs, which are completely uncorrelated for far apart point sources. However, because of ME, the point sources within a small ME region of the object plane will generate similar but shifted speckle patterns (i.e. PSFs) on the image plane. The mathematical representation of the image formation can be restated considering the ME as follows:
\begin{equation}
\label{eq2}
I(x,y)=\sum_R{\sum_{(i,j)\in R}{O(i,j)PSF_R (x-i,y-j)}} 
\end{equation}
where $R$ is the ME region having nearly identical $PSF_{ij}$ as $PSF_R$. Taking advantage of the shift-invariant property of $PSF_{ij}$, the superposition can be expressed as a convolution of different regions of the objects with their corresponding PSFs. 
\begin{equation}
\label{eq3} 
I_R= O_R*PSF_R=\sum_{(i,j)\in R}{O(i,j)PSF_R (x-i,y-j)}
\end{equation}
where $O_R$ is the portion of the object within the ME region, i.e. $O(i,j)$ for all $(i,j)\in R$. The object $O_R$ can be recovered by deconvolution using $PSF_R$  of the optical system, if we know its speckle pattern $I_R$. The deconvolution process is ideally expressed as follows. 
\begin{equation}
\label{eq4} 
O_R=deconv(I_R,PSF_R )=FFT^{-1} \left(\frac{FFT(I_R )FFT(PSF_R)^c}{|FFT(PSF_R)|^2}\right)    
\end{equation}
where  $(.)^c$ is the complex conjugate, $FFT(.)$ and $FFT^{-1} (.)$ are the Fourier transform and its inverse, respectively. The deconvolution in equation (\ref{eq4}) is possible because the convolution in spatial domain become the multiplication in Fourier domain. 
\begin{equation}
\label{eq5} 
FFT(O_R*PSF_E )=FFT(O_R )FFT(PSF_R ) 
\end{equation}
This deconvolution technique has been demonstrated to recover high resolution image of object within ME region of scattering media\cite{Zhuan16,Edrei16r}. Here for object $O$ extended over the ME region, its speckle pattern $I$ of the object is a composite response of the scattering light coming from the various regions of the object. Thus, the captured image can be expressed as follows:
\begin{equation}
\label{eq6} 
I=\sum_R{I_R} =\sum_R{O_R*PSF_R}
\end{equation}
Because of the random structure of the scattering media, spatially separated point sources beyond the ME region produce uncorrelated speckle patterns. This can be mathematically expressed as follows:
\begin{equation}
\label{eq7} 
PSF_{R1}\star PSF_{R2} = 
\begin{cases}
0  & \quad \text{if } {R1\neq R2 }\\
\delta  & \quad \text{if } {R1=R2}\\
\end{cases} 
\end{equation}
where $\star$ is the correlation operator, and $\delta$ is the spatial impulse function. Because of the relation $FFT(A\star B)= FFT(A)FFT(B)^c$, the following relationship can be deduced:
\begin{equation}
\label{eq8} 
FFT(I)FFT(PSF_R)^c=FFT(I_R )FFT(PSF_R)^c  
\end{equation}
Therefore, multiple region R of the object can be reconstructed from a single monochromatic image $I$ as below:
\begin{equation}
\label{eq9} 
O_R\approx deconv(I,PSF_R ) 
\end{equation}
In essence, each $PSF_R$ gives a limited FOV on the object plane, as seen in Fig. \ref{fig2}, where central FOV of the optical system is determined by central ME region and shown in the red dashed circular line. Actually, the scattering media might have different effective thickness for different angle illumination, therefore ME could slightly vary at different spatial position on the object plane, where spatial points have their own range of spatial shifted-invariance PSFs and unique spatial FOV. At the same time, all the spatial information of an object are multiplexed inside a single speckle image captured by the camera. One could use the unique regional PSF to reconstruct the spatial object $O_R$ for each region $R$, and obtain a full reconstruction of the large object by arranging the respective regions in space. 

\begin{figure}[htbp]
\centering\includegraphics[width=10cm]{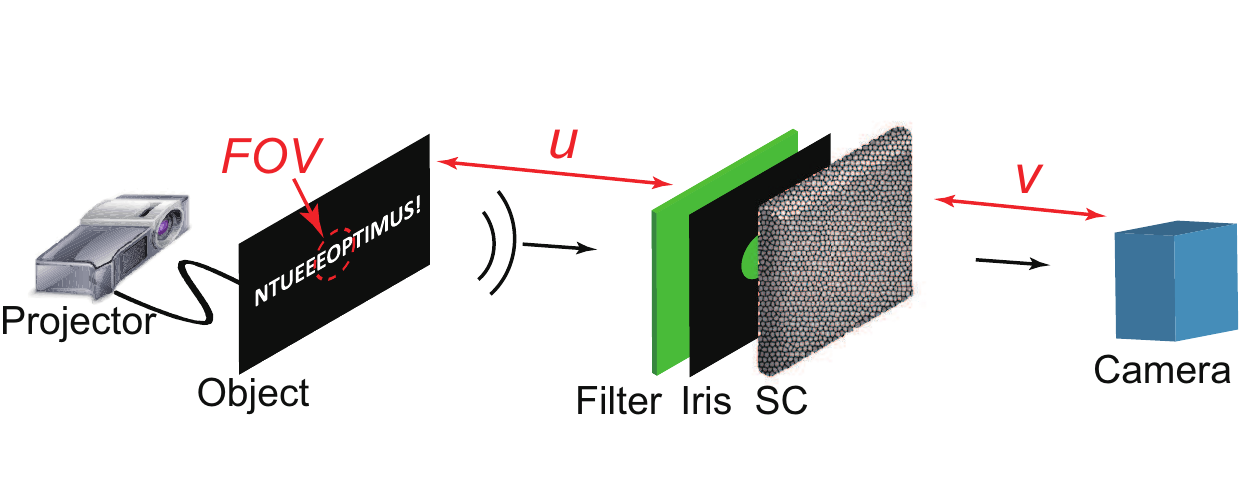}
\caption{Experimental setup. A common projector displays letters `NTU EEE OPTIMUS!' as an object for the imaging system. Light coming from the object passes through a green bandpass filter, an iris and the scattering medium, then generates a scrambled speckle pattern on a camera. Memory effect of the scattering medium gives a limited FOV on the object plane, as shown in the red dashed circle on the object plane. SC: scattering medium.}
\label{fig1} 
\end{figure}

\section{Experiment}
The schematic of our complete optical setup is presented in Fig. \ref{fig1}. A common projector creates a desired object as the input. In our experiment, the object is composed of a group of letters `NTU EEE OPTIMUS!'. Light coming from the object passes through a narrow bandpass green filter (center wavelength: $550\pm 2n\text{m}$, FHWM: $10\pm 2n\text{m}$), an iris and a thick scattering medium then generates a scattering speckle image on the camera (Andor Neo 5.5, $2560\times 2160$, pixel size $6.5 \mu\text{m}$). Iris with diameter $2 \text{mm}$ is used to remove the background light and control the intensity, aperture, speckle size and signal-to-noise ratio of the speckle pattern. The distance from the object to the scattering medium and that from the scattering medium to the camera are $u\approx170\text{mm}$ and $v\approx77.5\text{mm}$, respectively. Therefore the magnification of the setup is $M = v/u\approx0.456$. The thick scattering medium reduces the ME and therefore limits central FOV as shown in the dashed red circle in Fig. \ref{fig1}. Thus, only a partial object located within this FOV could be reconstructed with the central PSF through the deconvolution process. As the light from a point source (a single projector's pixel) going to the camera sensor is quite small, the exposure times for spatial PSF measurements are set as 10s; for the object with multiple point sources, the exposure times are about $0.03s\sim0.4s$. The applied Wiener deconvolution algorithm\cite{Gonzal08} is robust to the reconstruction noises and only takes about 0.5 second for each reconstruction with MATLAB on a normal PC (Intel Core i7, 16 GB memory). For each speckle patterns and PSFs, we divide them by their low frequency envelop to remove the hallow effect and sharpen the speckles. The resolution of the imaging system is defined by the speckle grain size, which depends on the numerical aperture of the system. Weiner deconvolution is performed by setting the noise level as the mean value of the deconvolving PSF.
\begin{figure}[htbp]
\centering\includegraphics[width=10cm]{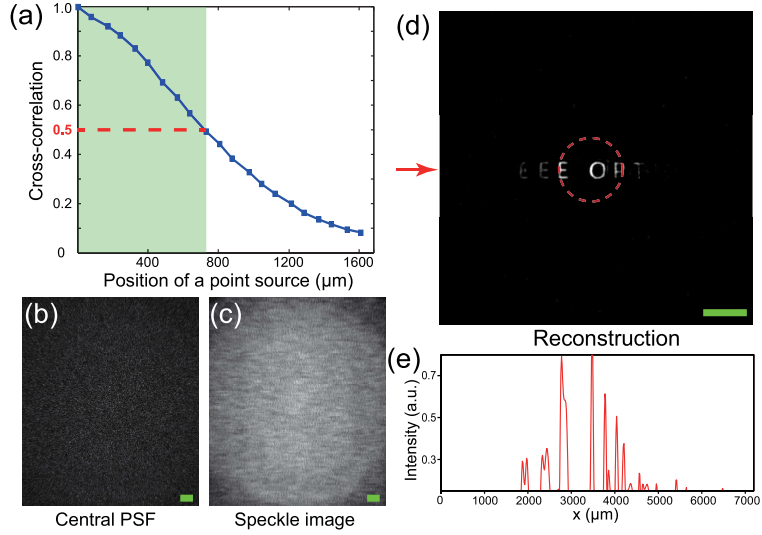}
\caption{Measurements and image recovery for the limited FOV in our optical system. (a) Cross-correlation coefficient between the central PSF and various spatial PSFs to identify the FOV on the object plane. Cyan region indicates a high-quality reconstructed area with cross-correlation coefficients larger than 0.5. (b, c) Central PSF and a captured speckle image of the object on the camera. (d) Reconstruction through a post-process algorithm indicates a successful recovery of the objects located in the dashed red circle, corresponding to the cyan region in (a). (e) Intensity distribution along the central horizontal direction.  Scale bars: $1000 \mu\text{m}$. The arrow indicate the central row of the image.}
\label{fig2} 
\end{figure}

Firstly, we need quantify and measure the central FOV in this setup. PSFs with respect to various positions of point sources on the object plane are measured as the projector successively lights up one pixel from the central position towards the corner along $x$ direction. The cross-correlation coefficients between the PSF at the center and PSFs at different positions are calculated to confirm the available central FOV according to the magnification between the object and image plane. Fig. \ref{fig2}a shows these cross-correlation coefficients as a function of the shifted distance on the object plane. We could see that only the object located in the cyan region could be reconstructed with high fidelity, if we define the FOV is the region with cross-correlation coefficient greater than 0.5. This ME region is corresponding to a strongly scattering medium with the effective thickness of 41.3 um. To demonstrate this limited FOV, we use the central PSF (Fig. \ref{fig2}b) and the speckle image (Fig. \ref{fig2}c) of the object in Fig. \ref{fig1} to do deconvolution. The reconstructed result and the intensity across the central horizontal direction are shown in Fig. \ref{fig2}d and Fig. \ref{fig2}e, where we clearly obtain the recovery of the object in the limited $720 \mu\text{m}$ radius FOV, same with the cyan region in Fig. \ref{fig2}a. For the reconstruction process, we remove the background with intensity less than 15\% of the maximum intensity for the clear display. The background noise comes from reconstruction artifact, which depends on the relationship between the actual signal of interest and the noise (in this work, noise are mainly from the uncorrelated speckle image generated by nearby regions of object). The full dimension of the speckle image and the PSF are used for the reconstruction, because the increase in number of speckles would reduce the reconstruction noise. However, the background noise is still unavoidable because the non-sparse object makes a poor speckle contrast within the dynamic range of camera. Beyond the FOV, partial recovery of the object could also be identified but the intensity are very dim due to the low cross-correlation coefficient (less than 0.5).

\begin{figure}[htbp]
\centering\includegraphics[width=13cm]{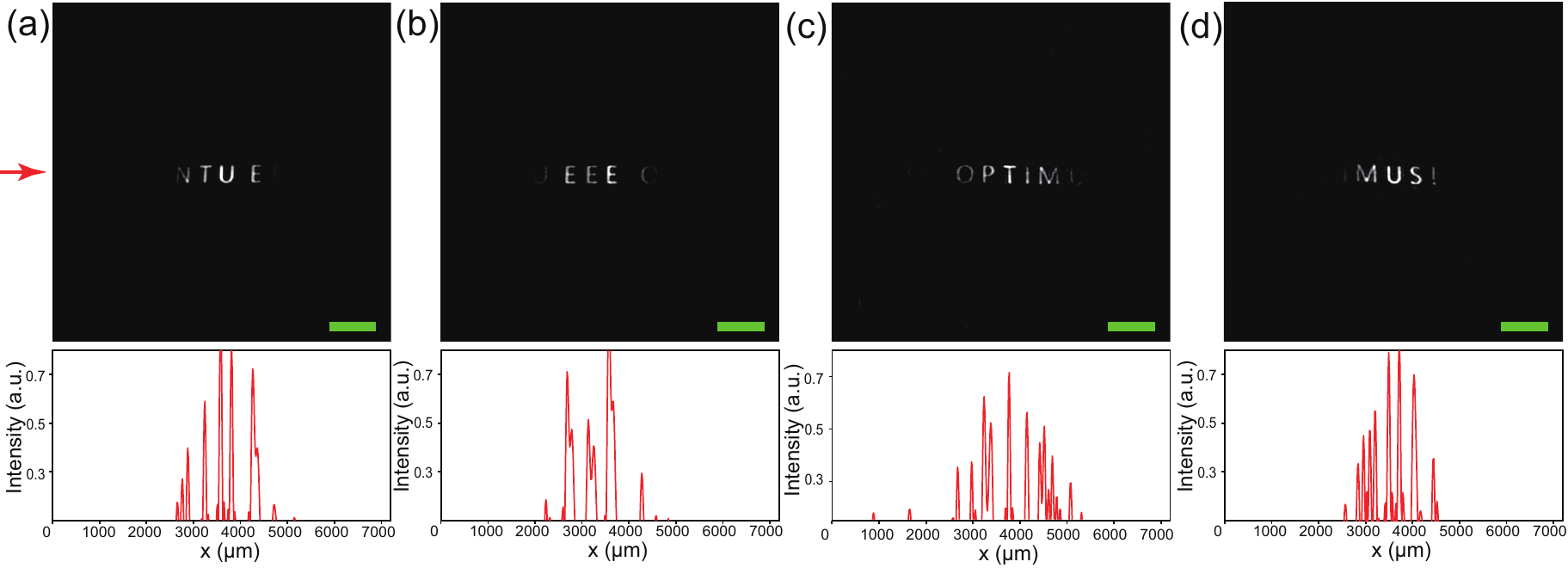}
\caption{Limited FOVs with various spatial PSFs. (a-d) A group of successful reconstructions of the object and corresponding 1D intensity distributions along central rows. Scale bars: $1000 \mu\text{m}$. The arrow indicate the central row of the image}
\label{fig3} 
\end{figure}

The above central PSF in Fig. \ref{fig2} only guarantees successful recovery image in the optical system with the linear shift-invariant property within a small region, where the deconvolution process reconstructs the object with high quality. Normally, the optical imaging system with scattering media is limited within the central FOV determined by the central PSF. However, the speckle image in Fig. \ref{fig2}c actually is the sum of the various PSFs from every point source on the object plane. In this work, we explore more information from the non-paraxial region and find an alternative way to extend the FOV while imaging with a strongly scattering medium using a single shot. There are multiple spatial FOVs satisfying the convolution operation in Equation (\ref{eq6}) for different spatial regions on the object plane. Equation (\ref{eq9}) shows that the information from various spatial regions could be recovered from the recorded single speckle image. We present various spatial point sources on the object plane and measure the corresponding PSFs, then utilize the same speckle image in Fig. \ref{fig2}c to execute deconvolution process. Fig. \ref{fig3} shows four typical reconstructions with four different spatial PSFs, the clear reconstruction centers and surrounding dim recoveries have same effects as Fig. \ref{fig2}d. At the bottom of each reconstructed image we have plotted 1D intensity of the pixel line running along center. We need note that the PSFs and available reconstruction regions (ME range) on image plane are unique and different for different spatial positions due to the random nature of the scattering medium. Therefore one could observe the different reconstruction FOV, for example, the FOV in Fig. \ref{fig3}c seems larger than that in Fig. \ref{fig3}a.
	
We have demonstrated our principle to demultiplexing the spatial information multiplexed into one single speckle image through various spatial PSFs in Fig. \ref{fig3}. However, for each individual reconstructed image, the information of interest are shifted to the center, and it is necessary to arrange them to form a single image with an enlarge FOV. We need to shift the reconstructed images according to the region associated with the PSFs, whose positions are predetermined. In fact, PSFs could record the shift of the speckle patterns and one could confirm their position through calculating the cross-correlation coefficient (even though the value maybe very low) to arrange the regional reconstructed images. If the regional PSF is measured for the point at position $(ux,uy)$, the recovered region on image plane should be centered at position $(vx,vy)$, where $v/u$ is the magnification. Now the challenge is to superpose these reconstructed regions to form the entire FOV. 

In reality, ME does not mean identical PSFs in the regional FOV. The cross-correlation coefficient between two PSFs reduces monotonically with their inter distance. In each FOV region, the cross-correlation coefficient of central PSF with other PSFs reduces when the point goes from the center to the edge (Fig. \ref{fig2}a); therefore, the intensity of recovered image decreases from center to the periphery. As presented above, we only need a single PSF for each regional FOV to present the function of a scattering lens. The position of the point source to take the PSF defines the center of lens and the PSF is the representative for this region. Any other point source on object plane also creates their own PSF whose cross-correlation coefficient with representative PSF represents the `transmission coefficient' of light from the point through the regional scattering lens in the deconvolution process. These transmission coefficients form a matrix $T$ for each scattering lens. The scattering medium is fixed; the cross-correlation coefficients for each region are not much different from the center regions. We assume that a single matrix $T$ can be used for all regions, and it can be written as follows: 
\begin{equation}
\label{eq10} 
O_R=O[\text{shift } R]T
\end{equation}
where $O[\text{shift }  R]$ is the object $O$ being shifted to center of the region $R$. Unfortunately both $O$ and $T$ are unknown and need to be recovered. Inspired by embedded pupil function recovery for Fourier ptychographic microscopy (EPRY-FPM)\cite{Zheng13,Ou14} we adopt an iterative update strategy to obtain both $O$ and $T$ as shown below.
\begin{equation}
\label{eq11} 
O^{(n+1)} [\text{shift }  R]=O^n [\text{shift }  R]+\alpha \frac{T^n}{\max(T^n )}\left(O_R - O^n [\text{shift }  R]T^n \right),   
\end{equation}
\begin{equation}
\label{eq12} 
T^{(n+1)}=T^n+\beta \frac{O^n}{max(O^n )}\left(O_R - O^n [\text{shift }  R]T^n \right)
\end{equation}

\begin{figure}[htbp]
\centering\includegraphics[width=12cm]{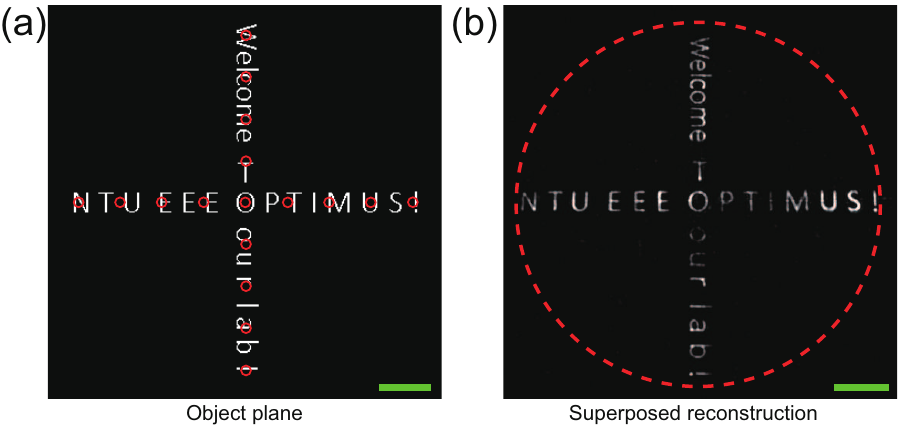}
\caption{Superposed reconstruction to enlarge the limited FOV. (a) Spatial distribution of the objects on the object plane. Red circles indicate the spatial positions of point sources for measuring the various spatial PSFs. (b) Superposed reconstruction image. Dashed red circle indicates the enlarged FOV. Scale bars: $1000 \mu\text{m}$.}
\label{fig4} 
\end{figure}

In the next experiment, the point sources used for the PSF measurement are shifted along both $x$ and $y$ direction with a fixed space. Therefore, we just need to retrieve the large FOV image from these reconstructed smaller FOV images in accordance with the shift in imaging plane using the aforementioned algorithm. Groups of letters `welcome TO our lab' and `NTU EEE OPTIMUS!' on vertical and horizontal axis are displayed as the object. The spatial PSFs for our optical system are from the point sources located at the various positions on the object plane, shown with red circles in Fig. \ref{fig4}a. A superposed reconstruction in Fig. \ref{fig4}b is $6650 \mu\text{m}$ width much larger than the reconstructed region by the central PSF alone $(2 \times 720 \mu\text{m})$. We have demonstrated the aim of the proposed technique: enlarging the FOV, which is done by utilizing the full capacity of the imager. The successful recovery will depend on the ability to resolve the contrast of the speckle image within the given dynamic range of the imager. The speckle contrast is reduced with increase in the illumination bandwidth, and having more bright areas in the object. Therefore, the reconstruction quality can be further enhanced with a narrow band illumination imaging and relatively sparser objects. 

\section{Conclusion}
We demonstrate an enlarged FOV for imaging with strongly scattering media by utilizing multiple ME regions. A single-shot speckle pattern contains essential spatial information of the large object. All the information is multiplexed on a single image randomly but pre-deterministically by the scattering medium. Any object region of interest can be retrieved by using the corresponding PSF of that region. The regional PSFs are fixed and only need to be recorded once for all time use. The deconvolution algorithm utilizes the shift invariance (i.e. spatial correlation) of the PSFs for image reconstruction, while the orthogonality between the regional PSFs (i.e. spatial decorrelation) makes them playing the role of spatial information demultiplexer. An iterative process for automatic weighted averaging is used to stitch the multiple regional images and form a large FOV image. The propose technique utilizes the imager to its full capacity in terms of both dimension and dynamic range to enlarge FOV imaging with strongly scattering media.

\section*{Funding}
Nanyang Technological University's start-up grant; Singapore Ministry of Education, MOE-AcRF Tier-1 grant (RG70/15); Singapore Ministry of Health's National Medical Research Council <CBRG-NIG (NMRC/BNIG/2039/2015)>.

\section*{Acknowledgments}
We would like to thank the financial supports from NTU start-up grant, Singapore MOE-AcRF Tier-1 grant (RG70/15) and the Singapore Ministry of Health's National Medical Research Council under its <CBRG-NIG (NMRC/BNIG/2039/2015)>.

\end{document}